\documentclass[12pt]{article}
\usepackage{natbib,graphicx,setspace,amsmath,amssymb,subfigure,url,showlabels}
\newtheorem{thm}{Theorem}
\newtheorem{lem}{Lemma}

\newtheorem{cor}{Corollary}

\begin{document}

\title{Semiparametric Bayesian Information Criterion for Model Selection in Ultra-high Dimensional Additive Models }         
\author{Heng Lian\\Division of Mathematical Sciences\\School of Physical and Mathematical Sciences\\Nanyang Technological University\\Singapore 637371\\Singapore}        
\maketitle
\begin{abstract}
For linear models with a diverging number of parameters, it has recently been shown that modified versions of Bayesian information criterion (BIC) can identify the true model consistently. However, in many cases there is little justification that the effects of the covariates are actually linear. Thus a semiparametric model such as the additive model studied here, is a viable alternative. We demonstrate that theoretical results on the consistency of BIC-type criterion can be extended to this more challenging situation, with dimension diverging exponentially fast with sample size. Besides, the noise assumptions are relaxed in our theoretical studies. These efforts significantly enlarge the applicability of the criterion to a more general class of models.

\textbf{Keywords:} Bayesian information criterion (BIC); Selection consistency; Sparsity; Ultra-high dimensional models; Variable selection.

\end{abstract}

\section{Introduction}  
With rapid increases in the production of large dimensional data by modern technology, more and more studies have focused on variable selection problems where the goal is to identify the few relevant predictors among a large collection of predictors, which might even outnumber the sample size due to the constraint of experimental costs. For example, in microarray experiments investigating genetic mechanisms of a certain disease, thousands of genes are assayed all at once while the number of samples is constrained by the cost of arrays as well as by the rarity of the disease in the population. 

In linear models with fixed dimension, performance of various criteria for variable selection is well known \citep{shao97}, including AIC \citep{akaike70}, BIC \citep{schwarz65}, $C_p$ \citep{mallows73} etc. In particular, BIC was shown to be consistent in variable selection. More recently, penalization approaches to variable selection have drawn increasing attention due to their stability and computational attractiveness \citep{tibshirani96,yuan06,fan01,zou06,liang2011}. Following this trend, \cite{wang07a} has shown that BIC computed along the solution path of the penalized estimator is also selection consistent. 

Nevertheless, these traditional criteria are too liberal for regression problems with high dimensional covariates, in that they tend to incorporate many spurious covariates in the model selected. On the positive side, modifications of BIC by using a statistically motivated larger penalty term can successfully address this problem, make the criterion provably consistent, and exhibit satisfactory performance in real applications \citep{wanglileng09,chenchen08}. Despite these efforts, the  works mentioned above, particularly the theoretical investigations, entirely focused on parametric linear models with Gaussian noise, while in many applications there is little a priori justification that the covariates actually have such simple linear effects on the responses. 

The additive model introduced by \cite{stone1985} represents a more flexible class of semiparametric models that allows a general transformation of each covariate to enter as an additive component. This raises an interesting question: is there an appropriately modified BIC-type criterion that can consistently identify the nonzero components in this class of semiparametric models? Although a similar question has been answered in an affirmative way in \cite{wang09} for fixed-dimensional varying-coefficient models, it remains a conjecture for high dimensional semiparametric problems. We note that \cite{huang10} has used modified BIC-type criterion in selecting the tuning parameter in group LASSO penalty for additive models, but they did not demonstrate the theoretical property of such a criterion. Compared to parametric models, the approximation errors for the component functions poses additional challenges to our analysis. 

In this paper, we will investigate the theoretical property of BIC-type criterion in additive models with the number of components $p$ growing much faster than sample size $n$. To be more specific, we assume $\log p=o(n^{2d/(2d+1)})$ where $d$ characterizes the smoothness (roughly the number of derivatives) of the component functions. Following the existing literature, we say the problem has a ultra-high dimensionality. On the other hand, the number of truly nonzero components is assumed to be fixed and does not diverge with sample size, for the same reason as discussed in \cite{huang10}. Besides, although we acknowledge that it might be restrictive to assume that all components have the same smoothness, it would be hard, if not impossible, to satisfactorily deal with the more general case. Finally, it is worth noting that we relax the Gaussian noise assumption used in \cite{chenchen08,wanglileng09} to sub-Gaussian noise. The Gaussian assumption was key to make the theoretical analysis tractable in those studies (see for example (B.3) in \cite{wanglileng09}). With sub-Gaussian noise, we need to resort to studying the tail probability of some quadratic forms involving sub-Gaussian random variables.

\section{Bayesian Information Criterion for Unpenalized Polynomial Spline Estimators}

Consider regression problems with observations $(Y_i, X_i), i=1,\ldots, n$ that are independent and identically distributed (i.i.d.) as $(Y,X)$, where $Y$ is a scalar response and $X=(X_1,\ldots, X_p)^T$ contains $p$ covariates. Substantial progress has been made on linear regression when $p$ is large, with or without penalty. 
Since fitting fully nonparametric models is infeasible for large dimensions, an elegant solution to relax the strong linearity assumption, known as the additive model \citep{stone1985,hastie90}, was proposed to avoid this difficulty, which is specified by
\begin{equation}\label{eqn:am}
 Y_i=\mu+\sum_{j=1}^p f_j(X_{ij})+\epsilon_i,
\end{equation}
where $\mu$ is the intercept, $f_j$ are unknown univariate component functions and $\epsilon_i$ are i.i.d. mean zero noises.

 Without loss of generality, we assume the distribution of $X_j$ is supported on $[0,1]$ and also impose the condition $E f_j(X_j)=0$ for identifiability. We use polynomial splines to approximate the components. Let $\tau_0=0<\tau_1<\cdots<\tau_{K'}<1=\tau_{K'+1}$ be a partition of $[0,1]$ into subintervals $[\tau_k,\tau_{k+1}),k=0,\ldots,K'$ with $K'$ internal knots. We only restrict our attention to equally spaced knots although data-driven choice can be considered such as putting knots at certain sample quantiles of the observed covariate values. A polynomial spline of order $q$ is a function whose restriction to each subinterval is a polynomial of degree $q-1$ and globally $q-2$ times continuously differentiable on $[0,1]$. The collection of splines with a fixed sequence of knots has a normalized B-spline basis $\{B_{1}(x),\ldots,B_{\tilde{K}}(x)\}$ with $\tilde{K}=K'+q$. Because of the centering constraint $E f_j(X_j)=0$, we instead focus on the subspace of spline functions $S^0_j:=\{s: s=\sum_{k=1}^{K}b_{jk}B_{jk}(x), \sum_{i=1}^n s(X_{ij})=0\}$ with basis $\{B_{jk}(x)=B_k(x)-\sum_{i=1}^n B_k(X_{ij})/n, k=1,\ldots, K=\tilde{K}-1\}$ (the subspace is $K=\tilde{K}-1$ dimensional due to the empirical version of the constraint). Using spline expansions, we can approximate the components by $f_j(x)\approx \sum_k b_{jk}B_{jk}(x)$. Note that it is possible to specify different $K$ for each component but we assume they are the same for simplicity (using the same $K$'s is reasonable when all components have the same smoothness parameter).

Suppose the true components are $f_{0j},1\le j\le p$, and the true intercept is denoted by $\mu_0$. We consider a sparse model where only the first $s$ components are nonzero. In unpenalized estimation, the following least squares estimation procedure is used to find the spline coefficients:
\begin{equation}\label{eqn:min}
(\hat{\mu},\hat{b})=\arg\min_{\mu,b}\sum_i(Y_i-\mu-\sum_{j=1}^p\sum_{k=1}^Kb_{jk}B_{jk}(X_{ij}))^2.
\end{equation}
However, the resulting estimator cannot be consistent when $p$ diverges at a sufficiently fast rate. Thus, we restrict our search on submodels where at most $M$ components are nonzero, where $M$ is a known fixed upper bound for $s$, and perform least squares regression with no more than $M$ components in (\ref{eqn:min}). Similar constraint is also imposed in \cite{chenchen08} for linear models.

Let
\[Z_j=\left(\begin{array}{cccc}
	B_{j1}(X_{1j})&B_{j2}(X_{1j})&\cdots&B_{jK}(X_{1j})\\
        \vdots&\vdots&\vdots&\vdots\\
	B_{j1}(X_{nj})&B_{j2}(X_{nj})&\cdots&B_{jK}(X_{nj})\\
	\end{array}\right)_{n\times K},\]
$Z=(Z_1,\ldots,Z_p)$, $Y=(Y_1,\ldots,Y_n)^T$. For any submodel indicated by $S\subseteq\{1,\ldots,p\}$, let $Z_S$ be the submatrix of $Z$ containing the columns in $S$, and similarly defined $b_S, \hat{b}_S$, etc. For notation convenience, we add $(1,\ldots,1)/\sqrt{K}$ as the first column of $Z, Z_S$ and define $a=(\sqrt{K}\mu,b^T)^T$, $a_S=(\sqrt{K}\mu,b_S^T)^T$, such that for the submodel $S$ (\ref{eqn:min}) can be written in matrix form as
\begin{equation}\label{eqn:min2}
\hat{a}_S=\min_{a_S}||Y-Z_Sa_S||^2.
\end{equation}
Let the true model be indicated by $S_0=\{1,\ldots,s\}$.

Now we can define the BIC-type criterion for the semiparametric model as
\begin{equation}\label{eqn:bic}
BIC(S)=\log(\|Y-Z_S\hat{a}_S\|^2)+|S|K\frac{\log n+\log p}{n},
\end{equation}
where $|S|$ is the size of the set $S$. The submodel $\hat{S}$ that achieves the minimum value of the above (over all submodels with $|S|\le M$) is chosen as the final model. The form of the above penalty is the same as that used in \cite{huang10} for group adaptive LASSO estimator, which is slightly different from that of \cite{chenchen08}, but easily seen to be asymptotically equivalent since $\log {p\choose j}\approx j\log p, j=1,\ldots,M$. The penalty in \cite{wanglileng09}, adapted to the semiparametric context here, is of the form $C_n|S|K\log n/n$ for some $C_n\rightarrow \infty$. We will try to be slightly more general and present our theoretical results for a general penalty term denoted by $pen(S)$.

The following technical conditions are assumed.
\begin{itemize}
\item[(c1)] The covariate vector $X$ has a continuous density supported on $[0,1]^p$. Furthermore, the marginal densities for $X_j, 1\le j\le p$ are all bounded from below and above by two fixed positive constants respectively.
\item[(c2)] The mean zero noises $\epsilon_{i}$ are independent of covariates, have variance $\sigma^2$, and are sub-Gaussian. That is there exists some $\alpha>0$ such that $E[\exp\{t\epsilon\}]\le \exp\{t^2\alpha^2/2\}$.
\item[(c3)] $f_{0j}, 1\le j\le s$ satisfies a Lipschitz condition of order $d>1/2$: $|f_{0j}^{(\lfloor d\rfloor)}(t)-f_{0j}^{(\lfloor d\rfloor)}(s)|\le C|s-t|^{d-\lfloor d\rfloor}$, where $\lfloor d \rfloor$ is the biggest integer strictly smaller than $d$ and $f_{0j}^{(\lfloor d\rfloor)}$ is the $\lfloor d\rfloor$-th derivative of $f_{0j}$. The order of the B-spline used satisfies $q\ge d+2$. 
\item[(c4)] The number of nonzero components is $s=O(1)$. 
\item[(c5)] $K\log(p n)/n\rightarrow 0$, $K\rightarrow\infty, K\log(p n)/n+K^{-2d}=o(\min_{1\le j\le s}\|f_{0j}\|^2)$, $pen(S_0)=o(\min_{1\le j\le s}\|f_{0j}\|^2)$, $K^{-2d}=o(pen(S)-pen(S_0))$ for $S\supsetneq S_0$, $K\log (p n)/n=O(pen(S)-pen(S_0))$ for $S\supsetneq S_0$.
\end{itemize}

Most of the assumptions are standard in the literature. Assumptions (c1)-(c4) are also assumed in \cite{huang10}. However, we will not assume that $\min_{1\le j\le s}\|f_{0j}\|$ is bounded away from zero as in assumption (A1) of \cite{huang10}. Instead, (c5) makes it clear that this quantity is allowed to converge to zero at a certain rate. Also note that in previous studies on the consistency of BIC-type criterion in linear models, Gaussian noise is assumed. We relax this assumption at the cost of more sophisticated arguments. We collect the assumptions on the convergence/divergence rate of different quantities in (c5). The expressions in (c5) can be simplified when $K\sim n^{1/(2d+1)}$ (this is the theoretically optimal choice of $K$ that balances bias and variance \citep{stone1985}) and $pen(S)=|S|K(\log n+\log p)/n$ (see Corollary \ref{cor:1} below).

\begin{thm}\label{thm:1}
Assume conditions (c1)-(c5). Then 
\[P(\hat{S}=S_0)\rightarrow 1.\]
\end{thm}

By this theorem, we know that with probability tending to 1, any model with size no larger than $M$ cannot be selected by BIC-type criterion, other than the true one. For particular form of the penalty function stated above, we have the following corollary.

\begin{cor}\label{cor:1}
If $K\sim n^{1/(2d+1)}$, $\log p=o(n^{2d/(2d+1)})$, $\min_{1\le j\le s}\|f_{0j}\|^2>>(\log (pn))n^{-2d/(2d+1)}$, then under conditions (c1)-(c4) the BIC-type criterion defined in (\ref{eqn:bic}) is selection consistent.
\end{cor}

\section{Bayesian Information Criterion for Penalized Estimators}
In the last section we stated that BIC-type criterion is consistent for variable selection for unpenalized estimators. However, even when the size of the submodels under consideration is constrained by $M$, brute-force search is still infeasible for large $p$. This is one of the reasons why penalized estimators become so popular in recent years. Here we briefly discuss how the results in the previous section can be extended to penalized estimator. 

In our context, the penalized estimator is defined by
\begin{equation}\label{eqn:pen}
\hat{a}_\lambda=\arg\min\|Y-Za\|^2+\sum_{j=1}^p p_\lambda(\|b_j\|),
\end{equation}
where $\lambda$ is the tuning parameter controlling the sparsity  of the solution, with larger $\lambda$ resulting in more components estimated as zero. Let $S_\lambda=\{j: \hat{b}_{\lambda j}\neq 0\}$ be the submodel represented by $\hat{a}_\lambda$. Here we focus on the group adaptive LASSO penalty since this is the one studied in \cite{huang10} for ultra-high dimensional additive models, although we expect selection consistency for estimators with SCAD penalty \citep{fan01} or MCP \citep{zhang2010nearly} can be derived in a similar way. Thus we assume all the conditions in \cite{huang10}. The BIC-type criterion for penalized estimator is defined as
\[BIC(\lambda)=\log(\|Y-Z\hat{a}_\lambda\|^2)+pen(S_\lambda),\]
and the optimal tuning parameter is $\hat{\lambda}=\arg\min_{\lambda>0}BIC(\lambda)$.

Following \cite{huang10}, for the group adaptive LASSO estimator, the penalty term in (\ref{eqn:pen}) is
$\sum_{j=1}^p \lambda\|b_j\|/\|\tilde{b}_j\|$ where $\|\tilde{b}_j\|$ is the initial group LASSO estimator. The following discussions are mainly extensions of arguments in \cite{wanglileng09}. Based on Corollary 2 in \cite{huang10}, if $K\sim n^{1/(2d+1)}$ and the tuning parameter is chose to be $\lambda_n\sim \sqrt{n}$, the estimator $\hat{a}_{\lambda_n}$ represents the correct model (that is $\hat{b}_{\lambda_n j}=0$ for $j>s$, or in other words $S_{\lambda_n}=S_0$). Since $\hat{b}_{\lambda_n j}=0$ for $j>s$, $\hat{a}_{\lambda_nS_0}=(\sqrt{K}\hat{\mu}_{\lambda_n},\hat{b}_{\lambda_n1},\ldots,\hat{b}_{\lambda_ns})^T$ must be the minimizer of 
\[\|Y-Z_{S_0}a\|^2+\sum_{j=1}^s \lambda_n\|b_j\|/\|\tilde{b}_j\|,\]
which yields by first order condition $\hat{a}_{\lambda_nS_0}=(Z_{S_0}^TZ_{S_0})^{-1}(Z_{S_0}^TY+\nu)$, where 
\[\nu=\partial \left.\sum_{j=1}^sp_\lambda(\|b_j\|)/\partial a\right|_{a=\hat{a}_{\lambda_nS_0}}=\lambda_n(0,\frac{\hat{a}_{\lambda_n1}^T}{\|\tilde{a}_1\|\cdot\|\hat{a}_{\lambda_n1}\|},\ldots,\frac{\hat{a}_{\lambda_ns}^T}{\|\tilde{a
}_s\|\cdot\|\hat{a}_{\lambda_ns}\|})^T.\]
We have $\|\nu\|^2=O(\lambda_n^2/K)=O(n/K)$. Thus
\begin{eqnarray*}
&&\|Y-Z\hat{a}_{\lambda_n}\|^2-\|Y-Z_{S_{\lambda_n}}\hat{a}_{S_{\lambda_n}}\|\\
&=&\|Z_{S_0}(Z_{S_0}^TZ_{S_0})^{-1}\nu\|^2-2(Y-P_{S_0}Y)(Z_{S_0}Z_{S_0}^TZ_{S_0})^{-1}\nu\\
&=&O((K/n)\|\nu\|^2+\sqrt{K+n/K^{2d}}\sqrt{K/n}\|\nu\|)\\
&=&O(\sqrt{K+n/K^{2d}}).
\end{eqnarray*}

Thus
\begin{eqnarray*}
&&BIC(\lambda)-BIC(\lambda_n)\\
&=&\log(\|Y-Z\hat{a}_\lambda\|^2)-\log(\|Y-Z\hat{a}_{\lambda_n}\|^2)+pen(S_\lambda)-pen(S_{\lambda_n})\\
&\ge&\log(\|Y-Z\hat{a}_{S_\lambda}\|^2)-\log(\|Y-Z\hat{a}_{\lambda_n}\|^2)+pen(S_\lambda)-pen(S_{\lambda_n})\\
&=&\log(\|Y-Z\hat{a}_{S_\lambda}\|^2)-\log(\|Y-Z\hat{a}_{S_{\lambda_n}}\|^2)+pen(S_\lambda)-pen(S_{\lambda_n})+O(\sqrt{K+n/K^{2d}})\\
&=&BIC(S_\lambda)-BIC(S_0)+O(\sqrt{K+n/K^{2d}}).
\end{eqnarray*}
A look at the proof for Theorem \ref{thm:1} in the Appendix shows that when $S_\lambda\neq S_0$ the gap between $BIC(S_\lambda)$ and $BIC(S_0)$ is actually larger than $O(\sqrt{K+n/K^{2d}})$, so the $O(\sqrt{K+n/K^{2d}})$ actually does not affect the result and we still have $BIC(\lambda)-BIC(\lambda_n)>0$ with probability tending to 1  uniformly over all $\lambda$ such that $S_\lambda\neq S_0$ and $|S_\lambda|\le M$.

\section{Conclusion and Discussion}
In this paper, we showed that the BIC-type criterion can be used in additive models with ultra-high feature dimensions to consistently select the true model. This paper is mainly of theoretical interest, and numerical evidence of its performance was contained already in \cite{huang10}. Although the BIC-type criterion is consistent for both unpenalized and the penalized estimators, computational constraints imply that the latter should be used in practice to avoid brute-force search over submodels. When the dimension of the feature space is so high that penalized approaches cannot be directly applied due to computational reasons, nonparametric independence screening procedure \citep{fan2009nonparametric} can be used as a first step to reduce the dimensionality.

The BIC-type criterion for penalized estimator focuses on the choice of tuning parameter $\lambda$ and ignores the choice of $K$ (the number of knots in B-spline approximation). In practice, $K$ can be fixed to a reasonable integral value as done in \cite{yu2002penalized,huang10,fan2009nonparametric} and some sensitivity analysis might be justified. It remains an open problem whether some criterion exists for data-driven choice of $K$ in high-dimensional contexts that has the desired theoretical property (in particular results in $K\sim n^{1/(2d+1)}$).

\section*{Appendix: Proofs}
By well-known properties of B-splines, there exists $b_{0j}=(b_{0j1},\ldots,b_{0jK})^T$ that satisfies the approximation property $\|\sum_kb_{0jk}B_{jk}(x)-f_{0j}(x)\|_\infty=O(K^{-d})$. Let $a_0=(\sqrt{K}\mu_0,b_{01}^T,\ldots,b_{0p}^T)^T$ and similarly define $a_{0S}$ for a submodel $S$. In our proofs, $C$ denotes a generic positive constant. We first present a Lemma which will be useful in the proof of the Theorem.

\begin{lem}\label{lem:1}
\[\sup_{S\supseteq S_0: |S|\le M} \left|\|Y-Z_S\hat{a}_S\|^2-\|Y-Z_Sa_{0S}\|^2\right|=O(nK^{-2d})+o(K\log(p n)).\]
\end{lem}
\textit{Proof of Lemma \ref{lem:1}}. We have
\begin{eqnarray}\label{eqn:main4}
&&\|Y-Z_S\hat{a}_S\|^2-\|Y-Z_{S}a_{0S}\|^2\nonumber\\
&=&-2(Y-Z_Sa_{0S})Z_S(\hat{a}_S-{a}_{0S})+\|Z_S(\hat{a}_S-a_{0S})\|^2\nonumber\\
&=&-2\epsilon^TZ_S(\hat{a}_S-a_{0S})-2(f_0(X)-Z_Sa_{0S})^TZ_{S}(\hat{a}_S-a_{0S})+\|Z_S(\hat{a}_S-a_{0S})\|^2,\nonumber\\
\end{eqnarray}
where $\epsilon=(\epsilon_1,\ldots,\epsilon_n)^T$ and $f_0(X)=(f_0(X_1),\ldots,f_0(X_n))^T$ with $f_0(X_i)=\mu_0+\sum_{j=1}^s f_{0j}(X_{ij})$ being the true regression function evaluation at covariate $X_i$.

By definition we have $\hat{a}_S=(Z_S^TZ_S)^{-1}Z_S^T(Z_S^Ta_{0S}+(f_0(X)-Z_S^Ta_{0S})+\epsilon)$ and thus $\hat{a}_S-a_{0S}=(Z_S^TZ_S)^{-1}Z_S^T(f_0(X)-Z_S^Ta_{0S})+(Z_S^TZ_S)^{-1}Z_S^T\epsilon$. Plugging this expression into (\ref{eqn:main4}) we get
\begin{eqnarray}\label{eqn:mainlem}
&&\|Y-Z_S\hat{a}_S\|^2-\|Y-Z_{S}a_{0S}\|^2\nonumber\\
&=&-2\epsilon^TP_S\epsilon-4\epsilon^TP_S(f_0(X)-Z_S^Ta_{0S})\nonumber\\
&&-2(f_0(X)-Z_S^Ta_{0S})^TP_S(f_0(X)-Z_S^Ta_{0S})+\|P_S\epsilon+P_S(f_0(X)-Za_{0S})\|^2\nonumber\\
&=&O(\epsilon^TP_S\epsilon+(f_0(X)-Z_S^Ta_{0S})^TP_S(f_0(X)-Z_S^Ta_{0S})),
\end{eqnarray}
where $P_S=Z_S(Z_S^TZ_S)^{-1}Z_S^T$ is a projection matrix. 

Obviously $(f_0(X)-Z_S^Ta_{0S})^TP_S(f_0(X)-Z_S^Ta_{0S})=O(nK^{-2d})$.
Next we will show $\sup_{S:|S|\le M}\epsilon^T P_{S}\epsilon=o(K\log(p n))$. Since we do not assume the errors are Gaussian, the quadratic form cannot be written as sum of chi-squared random variables. Fortunately we can still resort to results on quadratic forms for sub-Gaussian random variables. Specifically by Proposition 1.1 in \cite{mikosch1991estimates}, when $y\ge K\alpha^2$, we have 
\[P(\epsilon^TP_{S}\epsilon>\alpha^2MK+y))\le \exp\{-Cy/\alpha^2\},\]
and thus 
\[P(\sup_{S:|S|\le M}\epsilon^TP_{S}\epsilon>\alpha^2MK+y))\le O(p^M)\exp\{-Cy/\alpha^2\},\]
and if one takes $y=\delta K\log(p n)$ for any $\delta>0$, the above probability will tend to $0$. This shows $\sup_{S:|S|\le M}\epsilon^TP_{S}\epsilon=o(K\log(p n))$.\hfill $\Box$

\textbf{Proof of Theorem \ref{thm:1}.} The proof is split into two parts, considering the underfitted models (some nonzero components are not in $S$) and overfitted models (some zero components, as well as all nonzero components, are included in $S$) respectively.

\textit{Part 1: $S_0\not\subseteq S$. }

Let $\hat{a}_S$ and $\hat{a}_{S_0}$ be the least squares estimator under submodel $S$ and the true model $S_0$ respectively. Let $\tilde{S}=S\cup S_0$. With abuse of notation, $\hat{a}_S$ is also used to denote $|\tilde{S}|K+1$-dimensional vector where the coefficients not associated the submodel $S$ is filled in by zero. Similar statement applies to other notations such as $a_{S_0}$, $\hat{a}_{S_0}$ etc. Thus we can write expressions such as $Z_{\tilde{S}}\hat{a}_S$ even though $\tilde{S}\neq S$. That is, zero values are filled in to match the dimension. Then we have
\begin{eqnarray}\label{eqn:main}
&&\|Y-Z_{\tilde{S}}\hat{a}_S\|^2-\|Y-Z_{\tilde{S}}\hat{a}_{S_0}\|^2\nonumber\\
&=&-2(Y-Z_{\tilde{S}}\hat{a}_{S_0})^TZ_{\tilde{S}}(\hat{a}_S-\hat{a}_{S_0})+\|Z_{\tilde{S}}(\hat{a}_S-\hat{a}_{S_0})\|^2\nonumber\\
&=&-2\epsilon^TZ_{\tilde{S}}(\hat{a}_S-\hat{a}_{S_0})+2(Z_{\tilde{S}}^T\hat{a}_{S_0}-f_0(X))^TZ_{\tilde{S}}(\hat{a}_S-\hat{a}_{S_0})+\|Z_{\tilde{S}}(\hat{a}_S-\hat{a}_{S_0})\|^2.\nonumber\\
\end{eqnarray}

By existing results on spline estimator in additive models \citep{stone1985}, we know that when the true model is known, $\|\hat{a}_{S_0}-a_{0S_0}\|=O(K/\sqrt{n}+K^{-d+1/2})$. Besides, since some nonzero components in $a_{S_0}$ is estimated as zero in $\hat{a}_S$, we know $\|\hat{a}_S-{a}_{0S_0}\|\ge\min_{1\le j\le s}\|b_{0j}\|\ge C\sqrt{K}(\min_{1\le j\le s}\|f_{0j}\|-K^{-d})$ by the approximate property of splines. Thus uniformly for all $S\not\supseteq S_0$, 
\[\|\hat{a}_S-\hat{a}_{S_0}\|\ge\|\hat{a}_S-{a}_{0S_0}\|-\|{a}_{0S_0}-\hat{a}_{S_0}\|\ge C(\sqrt{K}\min_{1\le j\le s}\|f_{0j}\|-K/\sqrt{n}-K^{-d+1/2}).\]
Denote the right hand side above by $\gamma_n$, then the third term in (\ref{eqn:main}) is bounded below by $C(n/K)\gamma_n^2$ by Lemma 3 in \cite{huang10}. The absolute value of the second term is bounded by $\sqrt{nK^{-2d}}\sqrt{n/K}\gamma_n$ and thus is of smaller order than the third term. Finally we bound the first term in (\ref{eqn:main}) by 
\[-2\epsilon^TZ_{\tilde{S}}(\hat{a}_S-\hat{a}_{S_0})\ge -4\epsilon^T P_{\tilde{S}}\epsilon-\frac{1}{4}\|Z_{\tilde{S}}(\hat{a}_S-\hat{a}_{S_0})\|^2.\]
In the proof of Lemma \ref{lem:1} we showed that $\sup_{S:|S|\le M}\epsilon^T P_{\tilde{S}}\epsilon=o(K\log (p n))$ and thus by condition (c5), (\ref{eqn:main}) is bounded below a positive number at least as large as $C(n/K)\gamma_n^2$.
We have
\begin{eqnarray*}
&&BIC(S)-BIC(S_0)\\
&=&\log \left(1+\frac{\|Y-Z_S\hat{a}_S\|^2/n-\|Y-Z_{S_0}\hat{a}_{S_0}\|^2/n}{\|Y-Z_{S_0}\hat{a}_{S_0}\|^2/n}\right)+pen(S)-pen(S_0).
\end{eqnarray*}
Lemma \ref{lem:1} implies that $\|Y-Z_{S_0}\hat{a}_{S_0}\|^2/n\ge \|Y-Z_{S_0}a_{0S_0}\|^2/n-O(K^{-2d})-o((K/n)\log (p n))\ge \|\epsilon\|^2/(2n)-\|Z_{S_0}a_{0S_0}-f_0(X)\|^2/n-O(K^{-2d})-o((K/n)\log (p n))\rightarrow \sigma^2/2$. 
Thus 
\begin{eqnarray*}
&&BIC(S)-BIC(S_0)\\
&\ge&C(\min_{1\le j\le s}\|f_{0j}\|^2-K/n-K^{-2d})+pen(S)-pen(S_0),
\end{eqnarray*}
which is positive with probability tending to 1 by (c5). Thus $P(\min_{S\not\supseteq S_0:|S|\le M} BIC(S)-BIC(S_0)>0)\rightarrow 1$.


\textit{Part 2: $S\supsetneq S_0$.}

Lemma \ref{lem:1} showed that 
%
\begin{equation}\label{eqn:lem}
\sup_{S\supseteq S_0}\|Y-Z_Sa_{0S}\|^2-\|Y-Z_S\hat{a}_S\|^2=O(nK^{-2d})+o(K\log(p n)),
\end{equation}
and noting that $Z_Sa_{0S}=Z_{S_0}a_{0S_0}$ for $S_0\subseteq S$, we have
\begin{eqnarray*}
&&BIC(S_0)-BIC(S)\\
&=&\log\left(\frac{\|Y-Z_{S_0}\hat{a}_{S_0}\|^2}{\|Y-Z_S\hat{a}_S\|^2}\right)-(pen(S)-pen(S_0))\\
&\le&\log\left(\frac{\|Y-Z_{S_0}a_{0S_0}\|^2}{\|Y-Z_S\hat{a}_S\|^2}\right)-(pen(S)-pen(S_0))\\
&=&\log\left(1+\frac{\|Y-Z_Sa_{0S}\|^2-\|Y-Z_S\hat{a}_S\|^2}{\|Y-Z_S\hat{a}_S\|^2}\right)-(pen(S)-pen(S_0)).\\
\end{eqnarray*}
Using (\ref{eqn:lem}) and similar to the arguments at the end of Part 1, $\|Y-Z_S\hat{a}_S\|^2/n$ is bounded away from zero uniformly in $S\supseteq S_0$. And thus $BIC(S_0)-BIC(S)\le O(K^{-2d})+o(K\log(p n)/n)-(pen(S)-pen(S_0))<0$ with probability tending to 1.\hfill $\Box$

\bibliographystyle{jasa}
\bibliography{papers.txt,books.txt}

\end{document}